\documentclass[11pt]{article}
\usepackage{amsmath,epsfig,sint}
\font\sixrm=cmr6

\newcommand{\R}{\rm I\kern-.2emR}
\newcommand{\C}{\rm \kern.25em\vrule height1.4ex
depth-.12ex width.06em\kern-.31em C}
\newcommand{\N}{{\rm I\kern-.16em N}}
\newcommand{\Z}{{\rm Z\kern-.35em Z}}

\newcommand{\gr}{g_{{\hbox{\sixrm R}}}}
\newcommand{\Mr}{M_{{\hbox{\sixrm R}}}}

\begin{document}
\begin{titlepage}
\begin{flushright}
   AZPH-TH-99-03\\
   BUTP-99-04\\
   MPI-PhT/99-09\\
   March 1999
\end{flushright}

%\vskip 0.20 true cm
\begin{center}
{\Large\bf Comparison of the O(3) Bootstrap $\sigma$-Model\\[1.5ex]
 with the Lattice Regularization at Low Energies}
\end{center}
\vskip 0.25 true cm
\centerline{\large J\'anos Balog}
\vskip1ex
\centerline{Research Institute for Particle and Nuclear Physics}
\centerline{H-1525 Budapest 114, P.O.B. 49, Hungary}
\vskip 0.10 true cm
\centerline{\large Max Niedermaier}
\vskip1ex
\centerline{Department of Physics, University of Pittsburgh}
\centerline{Pennsylvania 15260, U.S.A.}
\vskip 0.10 true cm
\centerline{\large Ferenc Niedermayer\footnote{On leave from the
Institute of Theoretical Physics, E\"otv\"os University, Budapest,
Hungary}}
\vskip1ex
\centerline{Institute for Theoretical Physics, University of Bern}
\centerline{Sidlerstrasse 5, CH-3012 Bern, Switzerland}
\vskip 0.10 true cm
\centerline{\large Adrian Patrascioiu}
\vskip1ex
\centerline{Physics Department, University of Arizona}
\centerline{Tucson, AZ 85721, U.S.A.}
\vskip 0.10 true cm
\centerline{\large Erhard Seiler and Peter Weisz}
\vskip1ex
\centerline{Max-Planck-Institut f\"ur Physik}
\centerline{F\"ohringer Ring 6, D-80805 M\"unchen, Germany}
\vskip 0.20 true cm
\centerline{\bf Abstract}
\vskip 1.0ex
The renormalized coupling $\gr$ defined through the
connected 4-point function at zero external momentum
in the non-linear O(3) sigma-model in two dimensions, 
is computed in the continuum form factor bootstrap approach
with estimated error $\sim 0.3\%$.
New high precision data are presented for 
$\gr$ in the lattice regularized theory with standard
action for nearly thermodynamic lattices $L/\xi\sim 7$ and
correlation lengths $\xi$ up to $\sim 122$ and with the fixed 
point action for correlation lengths up to $\sim 12$.
The agreement between the form factor and lattice results is 
within $\sim 1\%$. 
We also recompute the phase shifts at low energy
by measuring the two-particle energies at finite volume,
a task which was previously performed by L\"uscher and Wolff using
the standard action, but this time using the fixed point action.
Excellent agreement with the Zamolodchikov S-matrix is found.

\vfill
\eject

\end{titlepage}

\section{Introduction}

The presently only known practical way to define a relativistic 
quantum field theory non-perturbatively 
in 4 dimensions is by using the lattice regularization. 
For example it is hoped that one will, 
once sufficiently powerful computers are available,
be able to answer the question whether QCD is the correct theory
of the strong interactions by studying the continuum limit
of the lattice theory. 

%It is however notoriously difficult to control the continuum limit 
%of a lattice theory. 
%
%
%Symanzik~\cite{symanzik} has shown that
%to all orders of perturbation theory ratios of physical
%quantities (of the same engineering dimension) approach
%their continuum limit as integer powers in the lattice spacing
%(up to logarithmic corrections). Unfortunately however
%there is no non-perturbative proof of this behavior. 
%All numerical simulation experiments have to resort to invoking 
%power-law approach as an implicit working hypothesis,
%and the integer nature of the power is intimately
%connected with the expected property of asymptotic freedom. 
%
%
%Symanzik \cite{symanzik} has shown that in the $\phi^4$ model in 4D
%order by order in renormalized
%perturbation theory, ratios of physical quantities
%(of the same engineering dimension) approach their continuum
%limit as integer powers in the lattice spacing (up to logarithmic
%corrections). Since it is not known that such behavior controls the 
%approach to the continuum limit in other models, especially outside of
%perturbation theory, the invocation of such
%a power-law approach to extrapolate data produced in numerical
%simulation experiments has merely the status of a (plausible)
%working  hypothesis
%and the integer nature of the power is intimately
%connected with the expected property of asymptotic freedom. 
%Here again the question whether
%the continuum limit of the lattice theory really describes an
%asymptotically free theory at high energies is highly non-trivial. 

It is however notoriously difficult to control 
the continuum limit of a lattice theory.
Analyses of lattice Feynman graphs, as initially 
performed by Symanzik \cite{symanzik}, show that
order by order in renormalized perturbation theory, 
physical quantities
approach their continuum limit as integer powers 
in the lattice spacing (up to logarithmic corrections). 
However, since it is not known that such behavior controls the 
approach to the continuum limit in the full non-perturbatively
defined theory, the invocation of such
a power-law approach to extrapolate data produced in numerical
simulation experiments has merely the status of a (plausible)
working  hypothesis. The integer nature of the powers
adopted in studies of theories such as QCD,
is considered to be connected to
the widely expected property of asymptotic freedom.
Here again, the very question whether the continuum
limit of the lattice theory really describes an
asymptotically free theory is highly 
non-trivial.\footnote{Indeed the authors of 
this paper are divided into two subsets
having different opinions on the probable answer!} 

This work is part of an on-going effort of 
the present authors to test whether the ``conventional wisdom"
is correct in a simpler model, the non-linear O(3) 
sigma model in 2 dimensions.
This model is, like QCD, perturbatively asymptotically free 
and also has instanton and superinstanton \cite{SUPIN} solutions. 
It has however 
classically the additional beautiful property of being integrable;
in particular there exist an infinite set of non-local conserved charges.
Assuming that the quantum theory has a mass gap and
the spectrum contains a vector multiplet of stable particles,
the existence of such conserved charges in the quantum theory
forbid particle production in scattering and, as shown  by
L\"{u}scher \cite{luscher} fixes the 2-particle S-matrix to that
previously postulated by Zamolodchikov and Zamolodchikov (ZZ) \cite{ZamI}
(up to CDD ambiguities).

All these properties were obtained starting from a formal Lagrangian,
where one first computes off-shell $N$-point functions
and goes on-shell via the LSZ formalism to obtain the S-matrix elements.
The so-called form factor bootstrap (FFB) approach 
\cite{smirnov,Karowski,KaWe}
proceeds in the other direction. One attempts to obtain
off-shell information starting from the knowledge (postulate) of the
stable particle spectrum and their S-matrix. In the first step
one constructs the form factors of (composite) operators,
satisfying all physical constraints 
(analyticity, generalized Watson theorem etc.) 
and then Green functions are obtained by saturating with
a complete set of states. This is a program
which is only feasible in a 
theory where there is no particle production
(i.e. only in 2 dimensions).
Even this program, for which there has
recently been a lot of progress \cite{JanosMaxHauer,JanosMax,BFKZ}, 
involves mammoth effort. 
 
Unfortunately since lattice regularization breaks these conservation laws
and the FFB relies on some nontrivial assumptions,
the expectation that the continuum limit of the
lattice O(3) model coincides with the FFB is not guaranteed. 
A first investigation
of this issue was by L\"{u}scher and Wolff \cite{lw} 
who computed the phase shifts on the lattice by measuring
two particle state energies in finite volume.
Their results (taking account of lattice artifacts) 
were completely consistent with the ZZ S-matrix.
In the course of a similar investigation of the nature of the
continuum limit of the O(2) model \cite{BNNPSWIII},
in order to test our programs  
and a slightly modified form of the analysis,
we repeated the measurement of the phase shifts in the O(3) model.
We also made simulations with a fixed point action \cite{FP},
and found indications that the lattice artifacts are
much smaller than in the case of the standard action.
An account of our investigations is given in sect.~4. 
Our results are again
in good agreement with the ZZ S-matrix \cite{ZamI}.

An off-shell quantity of physical interest 
is the current-current (J-J) correlation function. 
It has been shown that J-J  
computed in the FFB approach \cite{JanosMaxHauer,JanosMax} 
agrees well with
conventional renormalized perturbation theory at high energies, 
at least up to $p/M\sim10000$,
when the analytical value of the ratio 
of the mass $M$ to the $\Lambda$-parameter \cite{HMN} is used. 
Connected to this  
are two important inter-related properties;
firstly the ZZ S-matrix shows an ``on-shell form of AF",
in the sense that the phase shifts fall logarithmically
to zero with the energy at high energy (see sect.~4).
Secondly the thermodynamic Bethe ansatz, which is
used to compute $M/\Lambda$, reproduces the 2-loop
$\beta$-function coefficient.
The agreement between the FFB and lattice computations of  
J-J is also within $\sim 1\%$ 
for the entire range of momenta up to $\sim 40M$.

Despite this wealth of circumstantial evidence for
the validity of the conventional picture, there is still room
for doubt. In particular in a recent paper,
two of the present authors \cite{PS},
assuming a certain form of the lattice artifacts,  
have found statistically significant deviations
between the continuum limit of the lattice J-J correlation function 
and the FFB at low energies.

Unfortunately the J-J correlation function at low energies is a
quantity which behaves qualitatively similarly to 
that in a free theory. 
It is plausible that a difference between two theories 
would manifest itself more clearly in a quantity
which vanishes in the free theory e.g. the  
zero momentum coupling $\gr$ defined through the
connected 4-point function.
There is an enormous literature
on the computation of this quantity in the 2 (and higher)
dimensional non-linear sigma models, see e.g.  
ref.~\cite{peliss} and references therein. 
The main new contribution of
this paper, the computation of $\gr$ by the FFB,
is outlined in sect.~2.
This is the first time that this method has been used to compute a 
4-point function, and it is rather surprising that one is  
apparently able to get such a good approximation for $\gr$. 
%probably the most accurate one for~O(3)!

In sect.~3 we present results on $\gr$ using two different
lattice regularizations, the standard action (including new 
high precision data on thermodynamic lattices at large correlation lengths)
and the fixed point (FP) action. 
The nature of the approach to the continuum limit is
not so clear, but whatever (reasonable) extrapolation is made,
it agrees with the truncated FFB result better than $\sim 1\%$.

%\vfill
%\eject

%\input ./sect2
\section{Computation of $\gr$ in the continuum theory}

There have been various approximation schemes to compute low energy
(non-pertur-bative quantities) in the continuum formulation 
of the O$(n)$ models in two dimensions: 
the $g$--expansion \cite{FMPPT}, the
$\epsilon$-expansion \cite{peliss}, and the $1/n$ expansion
\cite{CPRVgR}. In this section we will present
a new approximation using the form factor bootstrap. 

\input./sect2

\subsection{$1/n$- and $\epsilon$-Expansions}

The leading order computations 
for the spectral integrals in the $1/n$-expansion 
have been performed in ref~\cite{CPRVU} 
\begin{gather}
\gamma_2=1+0.00671941 {1\over n} +O\left( {1\over n^2}\right)\,,
\label{1n1}\\
\delta_2=1+0.00026836 {1\over n} +O\left( {1\over n^2}\right)\,,
\label{1n2}\\
\nonumber
\end{gather}
\vskip -0.7truecm\noindent
and also for the coupling \cite{CPRVgR}
\begin{equation}
\gr={8\pi\over n}\left(1-0.602033{1\over n} +O({1\over n^2})\right),
\end{equation}
which gives the approximation $\gr=6.70$ for the case $n=3$.  

In the $g$-expansion one obtains $\gr=6.66(6)$ \cite{FMPPT}
and the $\epsilon$-expansion $\gr=6.55(8)$ \cite{peliss}.
Considering the rather short series in each case it 
is amazing how well the estimates by the various methods agree.

\subsection{Form factor bootstrap computation for $n=3$}

The form factor bootstrap aims at reconstructing $N$-point
functions of local operators of integrable field theories
from the knowledge of the spectrum of stable particles
and their S-matrix. A description can be found in Smirnov's
book~\cite{smirnov},
the review of Karowski~\cite{Karowski}, a recent paper
\cite{BFKZ} and in various articles of two 
of the present authors ~\cite{JanosMax}. 

To our knowledge this is the first time that the
method has been applied to the computation of 4-point functions.
The computation is rather involved and here we will only give a
very brief outline and present our results. The calculation
in this and in other integrable models will be 
described in detail in a forthcoming paper~\cite{BNNPSWII}.

We assume (as did the Zamolodchikov brothers in their 
construction of the S-matrix) that there are no bound states
in the O($n$) models. 
Then the spectral density $\rho$ has an expansion over contributions
from the intermediate states with an odd number of particles
(due to internal parity symmetry)
\begin{equation}
\rho(\mu)=\sum_{k=0}^{\infty} \rho^{(2k+1)}(\mu)\,,
\end{equation}  
and correspondingly the spectral integrals
\begin{gather}
\gamma_2=1+\sum_{k=1}^{\infty} \gamma_2^{(2k+1)}\,,
\\
\delta_2=1+\sum_{k=1}^{\infty} \delta_2^{(2k+1)}\,.
\end{gather}

The form factors are given by Smirnov~\cite{smirnov} for the O$(3)$ model
and have been recomputed in \cite{JanosMax}.
For example the matrix element of the Minkowski
operator $\hat{\sigma}^a(0)$ associated to the Euclidean field $\sigma^a$, 
connecting the vacuum to the 3-particle in-state
with rapidities 
$\theta_1\ge\theta_2\ge\theta_3$, is given by
\begin{equation}
\langle 0|\hat{\sigma}^a(0)
|a_1,\theta_1;a_2,\theta_2;a_3,\theta_3\rangle^{\rm
in}
=\sqrt{Z}{\cal F}^a_{a_1a_2a_3}(\theta_1,\theta_2,\theta_3)
\end{equation}
with
\begin{eqnarray}
&{\cal F}^a_{a_1a_2a_3}(\theta_1,\theta_2,\theta_3)
=\pi^3\prod_{i<j} \psi(\theta_i-\theta_j)
\nonumber
\\
\cr
&\cdot
\Bigl[ (\theta_3-\theta_2)\delta^a_{a_1}\delta_{a_2a_3}
+(\theta_1-\theta_3-2\pi i)\delta^a_{a_2}\delta_{a_3a_1}
+(\theta_2-\theta_1)\delta^a_{a_3}\delta_{a_1a_2}
\Bigr]\,,
\end{eqnarray}
where
\begin{equation}
\psi(\theta)={(\theta-\pi i)\over \theta(2\pi i-\theta)}
\tanh^2{\theta\over2}\,.
\end{equation}
The expression of the 5-particle matrix element 
is also known explicitly but it is much too
long to be written here.
Using these results the 3- and  5-particle
contributions to $\gamma_2$ and $\delta_2$ are given in 
table~\ref{spectralmoments}. It seems that the series
converge extremely rapidly and we would estimate
\begin{eqnarray}
\label{gamma2}
&\gamma_2=1.001687(1)\,,
\\
&\delta_2=1.000034657(1)\,,
\label{delta2}
\end{eqnarray}
where the estimated errors come from inspecting the pattern
of relative $n$-particle contributions suggested by the
1,3,5-particle states.

\begin{table}[t]
\centering
\begin{tabular}[t]{r|r|r}
\hline
$r$ &$\gamma_2^{(r)}\ \ \ \ \ \ \ \ $& 
$\delta_2^{(r)}\ \ \ \ \ \ \ \ $  \\[0.5ex]
\hline \hline    

3 & $1.67995(1) \times 10^{-3}$ & $3.46494(1) \times 10^{-5}$\\[0.5ex]

5 & $6.622(1)\times 10^{-6}$ & $7.114(1) \times 10^{-9}$\\[0.5ex]

\hline
\end{tabular}
\caption{\footnotesize $r$-particles contribution to $\gamma_2, \delta_2$}
\label{spectralmoments}
\end{table}

The 4-point function has an expansion in terms of 
contributions of intermediate states with $l,m,n$ particles respectively
%\begin{equation}
%\widetilde{S}^{a_1a_2a_3a_4}(k_1,k_2,k_3,k_4)=\sum_{{\rm perms}\,P}
%W^{a_{P1}a_{P2}a_{P3}a_{P4}}(k_{P1},k_{P2},k_{P3},k_{P4})
%\end{equation}
%where $k_i=(k_{i1},k_{i2})$,
%\begin{multline}
%W^{a_1a_2a_3a_4}(k_1,k_2,k_3,k_4)
%=(2\pi)^5\delta^{(2)}(k_1+k_2+k_3+k_4)
%\sum_{\underline{l},\underline{m},\underline{n}}^{}
%{\delta(P_{\underline{l}}+k_{11})\over
%E_{\underline{l}}-ik_{12}}
%\\
%\cdot
%{\delta(P_{\underline{m}}+k_{11}+k_{21})\over
%E_{\underline{m}}-ik_{12}-ik_{22}}
%{\delta(P_{\underline{n}}-k_{41})\over E_{\underline{n}}+ik_{42}}
%\langle 0 |\hat{\sigma}^{a_1}(0)|\underline{l}\rangle
%\langle \underline{l} |\hat{\sigma}^{a_2}(0)|\underline{m}\rangle
%\langle \underline{m} |\hat{\sigma}^{a_3}(0)|\underline{n}\rangle
%\langle \underline{n} |\hat{\sigma}^{a_4}(0)|0\rangle\,,
%\label{multline}
%\end{multline}
\begin{multline}
\widetilde{S}^{a_1a_2a_3a_4}(k_1,k_2,k_3,k_4)
=(2\pi)^2\delta^{(2)}(k_1+k_2+k_3+k_4)
\\
\cdot M^{-6}\sum_{{\rm perms}\,P}
V^{a_{P1}a_{P2}a_{P3}a_{P4}}(k_{P1},k_{P2},k_{P3},k_{P4})
\end{multline}
where $k_i=(k_{i1},k_{i2})$,
\begin{multline}
V^{a_1a_2a_3a_4}(k_1,k_2,k_3,k_4)=(2\pi)^3 M^6
\sum_{\underline{l},\underline{m},\underline{n}}^{}
{\delta(P_{\underline{l}}+k_{11})\over
E_{\underline{l}}-ik_{12}}
{\delta(P_{\underline{m}}+k_{11}+k_{21})\over
E_{\underline{m}}-ik_{12}-ik_{22}}
\\
\cdot
{\delta(P_{\underline{n}}-k_{41})\over E_{\underline{n}}+ik_{42}}
\langle 0 |\hat{\sigma}^{a_1}(0)|\underline{l}\rangle
\langle \underline{l} |\hat{\sigma}^{a_2}(0)|\underline{m}\rangle
\langle \underline{m} |\hat{\sigma}^{a_3}(0)|\underline{n}\rangle
\langle \underline{n} |\hat{\sigma}^{a_4}(0)|0\rangle\,,
\label{multline}
\end{multline}
where $\underline{l},\underline{n}$ run over all states with 
odd numbers $l,n$ of particles, 
and $\underline{m}$ over all states with an even number
$m\ge 0$.
The somewhat symbolic $\sum$ in (\ref{multline}) really means 
in addition to the summation over all internal quantum numbers 
of the particles and integration over all particle rapidities
a sum over the integers $l$, $m$ and $n$.
The limit of zero momenta is very delicate because each term 
in the sum is a distribution in the momenta where the singularities
occur when certain linear combinations of the momenta are zero.
In particular the contributions from terms in the above sum
with $m=0$, not only cancel the disconnected pieces
$\widetilde{S}-\widetilde{S}_c$,  but also produce
extra terms proportional to $\delta(k_{11}+k_{21})$.
The singularities must be canceled by other terms in the sum
with $m>0$ e.g. the singularity from the contribution $1$-$0$-$1$
is canceled by that of the contribution $1$-$2$-$1$.
We can avoid this problem if, as is assumed in the following,
we restrict ourselves to momenta where $k_{i1}+k_{j1}\ne 0$
for any $i\ne j$. 
An additional technical complication comes from the fact that 
each term in the sum is rather involved because the
matrix elements have parts with differing connectivity
properties, e.g. the matrix element occurring in the 1-2-1
contribution ($\theta_1>\theta_2$):
\begin{eqnarray}
&Z^{-\frac12}
\langle b,\phi|\hat{\sigma}^a(0)|a_1,\theta_1;a_2,\theta_2\rangle^{\rm in}
={\cal F}^a_{ba_1a_2}(\phi+i\pi-i\epsilon,\theta_1,\theta_2)
\nonumber
\\
\cr
&+4\pi\delta^a_{a_2}\delta_{ba_1}\delta(\phi-\theta_1)
+4\pi\delta(\phi-\theta_2)S_{a_1a_2;ab}(\phi-\theta_2)\,,
\end{eqnarray}
where the S-matrix elements $S_{a_1a_2;ab}$ are 
given in sect. 4.

The practicability of the computation of the
zero-momentum coupling using the form factor bootstrap
approach obviously depends crucially on the question whether
the sum over intermediate states for the 4-point function,
\begin{equation}
\gamma_4=\sum_{l,m,n} \gamma_{4;lmn}
\end{equation}
converges rapidly.
We started\footnote{We hope to be able to work out the $l+m+n=8$
contributions in the Ising model soon.} 
to investigate this question in the Ising model and we found that
the $l+m+n=6$ contributions are much smaller than the leading
$1$-$2$-$3$ term \cite{BNNPSWII}.
Fortunately, for the O(3) model under investigation here
the situation is rather similar.
The contributions of the $l$-$m$-$n$ intermediate states 
with $l+m+n\le 6$ to 
$\gamma_4$  are given in table~\ref{gamma4}. 

\begin{table}[t]
\centering
\begin{tabular}[t]{r|r}
\hline
$l,m,n$ &$\gamma_{4;lmn}$  \\[0.5ex]
\hline \hline

$1,2,1$ & $-4.16835(1)\phantom{5}$\\[0.5ex]

$1,2,3$ & $\phantom{-}0.05175(1)\phantom{5}$\\[0.5ex]

$3,2,1$ & $\phantom{-}0.05175(1)\phantom{5}$\\[0.5ex]

$1,4,1$ & $-0.004065(1)$\\[0.5ex]

\hline
\end{tabular}
\caption{\footnotesize $l$-$m$-$n$-particles contribution to $\gamma_4$}
\label{gamma4}
\end{table}

The leading $1$-$2$-$1$ contribution is a
factor $\sim 42$ greater in magnitude than the sum of
$l$-$m$-$n$ contributions with $l+m+n=6$. It is extremely
difficult to bound the rest of the
contributions, especially
since the signs are not known in general. Even the computation of
the states with $l+m+n=8$ would be quite an undertaking.
But assuming that the pattern in table~\ref{gamma4} continues
(as for the case of the 2-point function) and that the
sum of the remaining contributions $l+m+n\ge 8$ 
is $\le 10\%$ of the sum of the
$l+m+n=6$ contributions, we obtain the result
\begin{equation}
\gamma_4=-4.069(10)\,,
\end{equation}
and hence our final result
\begin{equation}
\label{grffb}
\gr=6.770(17)\,.
\end{equation} 

%\vfill
%\eject

\newcommand{\p}{{\phantom 0}}

\section{Lattice computations of $\gr$}

In the framework of the lattice regularization there are two methods
to compute $\gr$ in the O($n$) models. The first is using
high temperature (strong coupling) expansions
and the second by numerical simulations.

For numerical simulations we consider a square lattice with both
the standard action
\begin{equation}
S=-\beta\sum_{x,\mu}\sigma(x)\cdot\sigma(x+\hat{\mu})\,,
\end{equation}
where $\sigma(x)\cdot\sigma(x)=\sum_a\sigma^a(x)\sigma^a(x)=1$
and the fixed point (FP) action of ref. \cite{FP}.

\subsection{High temperature expansion}

Concerning the high temperature (HT) expansion for the standard action, 
long series \cite{BC} have been
obtained for the 2- and 4-point functions and for the second moment
$\mu_2$ defined by
\begin{equation}
\mu_2\delta^{a_1a_2}=\sum_x x^2
\langle\sigma^{a_1}(x)\sigma^{a_2}(0)\rangle 
\end{equation}
to obtain $\Mr$ through $\Mr^2=4G(0)/\mu_2$,
where $G(k)$ is defined analogously to eq.~(\ref{twopoint}) but
with the integral replaced by a sum.
The analyses of the HT expansion for the spectral moments 
give $\gamma_2=1.0013(2)$ \cite{CPRVU} and
$\delta_2=1.000029(5)$ \cite{rossi}. The agreement with the
FFB values eqs.(\ref{gamma2},\ref{delta2}) is acceptable;
note that these are smaller than that anticipated from
the leading order of the $1/n$ approximation, eqs. (\ref{1n1}), (\ref{1n2}).

The coupling is defined as the continuum limit
\begin{equation}
\gr=\lim_{\xi\to\infty}\gr(\xi)\,,
\end{equation}
where $\xi$ is the correlation length in lattice units.
The analysis is hampered by the lack of rigorous knowledge 
of the position of the critical point and the exact approach to it.
In particular, the conventional wisdom 
that the critical point is at $\beta=\infty$
is usually built into the analyses. The various Pad\'e approximations
show the coupling falling rapidly as $\beta$ increases in the 
region of small $\beta$, then a region of rather flat behavior
after which the various approximations show diverse behavior
(see e.g. fig.12 in ref.~\cite{CPRV}) making
error estimates rather difficult 
\footnote{A result is considered
more reliable if in the region where the coupling flattens
the approximants have no complex singularities near the real axis.
Unfortunately for the special case $n=3$ there do tend to be nearby 
singularities \cite{butera}.}.

In ref.~\cite{CPRVhT} Campostrini et al
quote for the case $n=3$ the result $\gr=6.6(1)$,
and in a more recent publication Pelissetto and Vicari
cite $6.56(4)$~\cite{peliss}. 
Butera and Comi on the other hand are rather cautious, and
did not quote a value for the case $n=3$ in ref.~\cite{BC};
if pressed they would at present cite $\gr=6.6(2)$ \cite{butera}.

\subsection{Numerical simulations}

Monte Carlo computations of $\gr$ have a long history,
see e.g. refs.~\cite{FMPPT,kim}. In order to attempt to
match the apparent precision attained in the FFB approach 
outlined in section 4, we decided to perform even more precise
measurements than were carried out previously.
 
We work here on a square lattice
of size $L$ in each direction and periodic boundary conditions.
The coupling in finite volume is defined
through Binder's cumulant
\begin{equation}
\gr(\xi)=\lim_{L\to\infty}\gr(\xi,L)\,,
\end{equation} 
\begin{equation}
\gr(\xi,L)=
\left({L\over\xi}\right)^2
\Bigl[1+{2\over n}-
{\langle (\Sigma^2)^2\rangle \over \langle \Sigma^2\rangle^2}\Bigr]
\end{equation}
where $\Sigma^a=\sum_x \sigma^a(x)$. In this definition 
we have taken 
(as in ref.~\cite{FMPPT}):
\begin{equation}
\xi={1\over 2 \sin(\pi/L)} \sqrt{{G(0)\over G(k_0)}-1}\,,
\label{xisecmom}
\end{equation}
where $k_0=(2\pi/L,0)$. 
%It differs from the `true' (exponential) 
%correlation length by less than 2 parts in 1000, if $L/\xi\geq 7$.
In this paper we will use $\xi$ to denote the second moment correlation
length, to which (\ref{xisecmom}) converges, for large $L$.
Although conceptually different, 
$\xi$ is very close to the exponential correlation length 
($\xi_{{\rm exp}}$):
also using (\ref{nxi1}) and (\ref{nxi2}) in the definition (\ref{xisecmom})
the FFB results (\ref{gamma2}) and (\ref{delta2}) yield
\begin{equation}
{\xi_{{\rm exp}}\over\xi}=\sqrt{{\gamma_2\over\delta_2}}=1.000826(1)\,.
\end{equation}

%%%%%%%%%%%%%%%%%%%%%%%%%%%%%%%%%%%%%%%%%%%%%%%%%%%%%%%%%%%%%%
% Here begins the passage by Adrian and Erhard
%
%
%The standard action
%Monte Carlo investigation of $\gr$ was performed on an 
%SGI 2000 computer. The measurements were performed using a method similar
%to the improved cluster estimator of~\cite{lw}.
%We took 5 measurements at correlation lengths
%ranging from 11 to 122 at $L/\xi \sim 7$. These measurements were used 
%to investigate the approach to the continuum in a box of physical size 
%$\sim 7\xi$. To study the finite volume effects, we repeated the runs at 
%$\xi \sim 11$ on two other lattices with $L/\xi\sim 5.5$ and $L/\xi\sim 13$. 
%The results of all these runs (together with the preliminary results
%corresponding to $\xi\sim167$, $230$ and $309$)
%are recorded in table~\ref{grdata}.
%
%
The standard action Monte Carlo measurements were performed  
using a method similar to the cluster estimator of~\cite{lw}.
We measured $\gr$ at correlation lengths
ranging from 11 to 122 at $L/\xi \sim 7$. These measurements were used 
to investigate the approach to the continuum. To study the finite 
volume effects, we repeated the runs at 
$\xi \sim 11$ on three other lattices with $L/\xi\sim5.5$,
$L/\xi\sim9$ and $L/\xi\sim13$. 
The results of all these runs (together with the preliminary results
corresponding to $\xi\sim167$, $230$ and $309$)
are recorded in table~\ref{grdata}.

%
%
%In this table we also indicate the number of measurements. These were 
%performed using the cluster algorithm as follows: one run consisted of 
%100,000 clusters used for thermalization, followed by 20,000 sweeps
%of the lattice used for measurements. Each run was repeated after 
%changing the initial configuration. One such run was considered as one 
%independent measurement. The error was computed from this sample by 
%using the jack-knife method.
%
%

In this table we also indicate the number of measurements. 
Each run consisted of 20k sweeps of the lattice with measurements
after each sweep. The error was computed from this sample by 
using the jack-knife method.

We have measured $\gr$ with the FP action at three different 
values of $\beta$: 0.70, 0.85 and 1.00, corresponding to 
correlation length $\xi\approx 3.2$, $6.0$ and $12.2$,
at the values of $z=L/\xi$ in the range $5.4 \ldots 8.2$.

%
%
%Our results suggest that for $L/\xi\sim 7$ 
%the coupling $\gr$ has reached its
%thermodynamic value within our precision, at least for $\xi\sim 11$.
%(See, however, Fig. \ref{fig_grfp}.)
%
%

To get a feeling of the finite volume effects we took
the expression for $\gr$ in the leading order $1/n$
expansion from \cite{CPRVgR} 
and simply replaced the integral over momenta by
a discrete sum. In this way we obtained
\begin{equation}
\gr[L]=\gr[\infty][1-a_1\sqrt{z}{\rm e}^{-z}(1+O(1/z))+\cdots]
\label{FS1n}
\end{equation}
with $a_1=\sqrt{8\pi}=5.013$ for large $z=L/\xi$. 
%The finite size effects in this case 
%are not so small, as anticipated in ref.~{FMPPT} 
%but for $z=7$ they are only $\sim 1\%$ 
%(and the sign of the effects is as indicated in the O(3) MC data).
%The sign of the finite size effects predicted by this expression
%is the same as indicated in the O(3) MC data and its magnitude is 
%$\sim1\%$ for $z=7$.

\begin{figure}[htb]
\begin{center}
%\vspace*{-1mm}
\epsfig{file=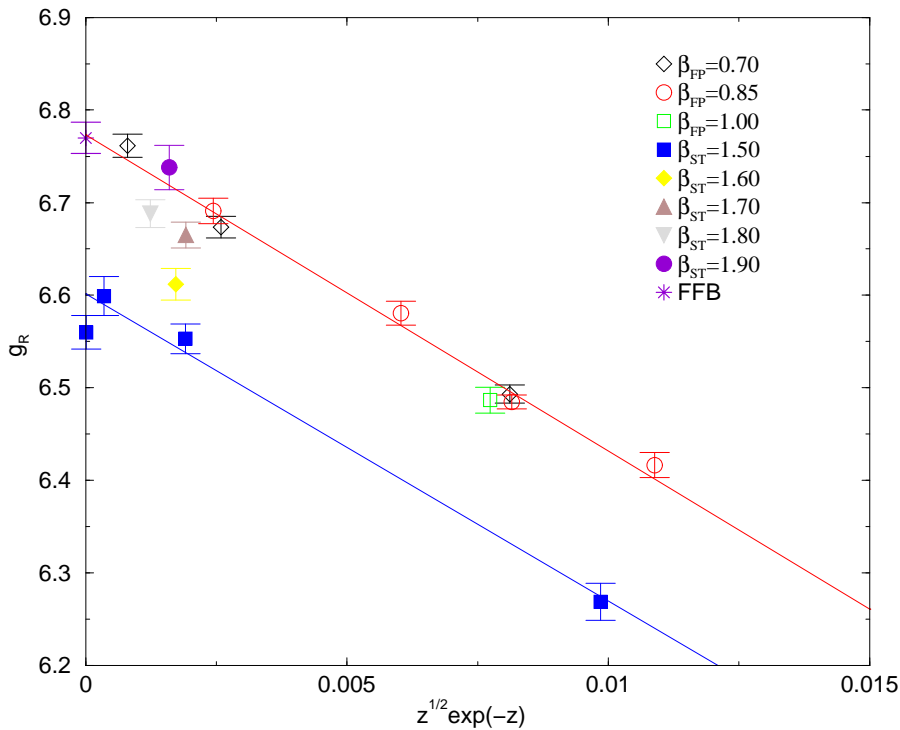,angle=-90,width=12cm}
%\vspace*{-10mm}
\end{center}
\caption{\footnotesize The coupling $\gr(z,\beta)$ for the FP (open symbols)
and standard actions (filled symbols) plotted vs. $\sqrt{z}\exp(-z)$.
The correlation length for the FP action are in the range
$3.2\ldots 12.2$ while for the standard action 
in $11 \ldots 122$. The theoretical value from the FFB is also shown.
The linear fits are motivated by the the $1/n$ prediction (\ref{FS1n})
and are based on the $\beta_{\rm ST}=1.50$ and $\beta_{\rm FP}=0.85$ 
data, respectively.}
\label{fig_grfp}
\end{figure}

Figure \ref{fig_grfp} shows these results for $\gr$ plotted against 
the combination $\sqrt{z}\exp(-z)$, 
motivated by the $1/n$ result (\ref{FS1n}).

%Martin L\"{u}scher has also considered general 
%finite volume effects on the coupling in the cylinder topology.
%Should we look at the finite volume effects in the torus topology
%in $1/n$ approximation?  

%The approach to the continuum is a more delicate issue. 
We have determined the MC prediction of $\gr$ both for the standard
action and the FP action.
%First we estimated the finite size effects as follows.
Making a linear fit in $\sqrt{z}\exp(-z)$ to the four data points
at $\beta=1.50$ for the standard action (see fig.~\ref{fig_grfp})
one obtains
\begin{equation}
\gr(z=\infty,\xi=11.0)=6.60(1)\,,
\label{neweq}
\end{equation}
with $a_1=5.0(4)$.
%As we see from fig. \ref{fig_grfp} the $1/n$ formula (\ref{FS1n})
%gives a very good description of the finite size effects for both the
%standard action and the FP action. Even the fitted value of the coefficient
%$a_1$ (see below) is close to that predicted by the $1/n$ expansion.
We therefore used (\ref{FS1n}) to \lq\lq renormalize" the data 
with $z\sim7$ in table \ref{grdata} to the common physical size
$z=7.25$.

\begin{figure}[t]
%\centerline{\epsfxsize=12.0cm\epsfbox{gr_xi7_l.eps}}
\epsfig{file=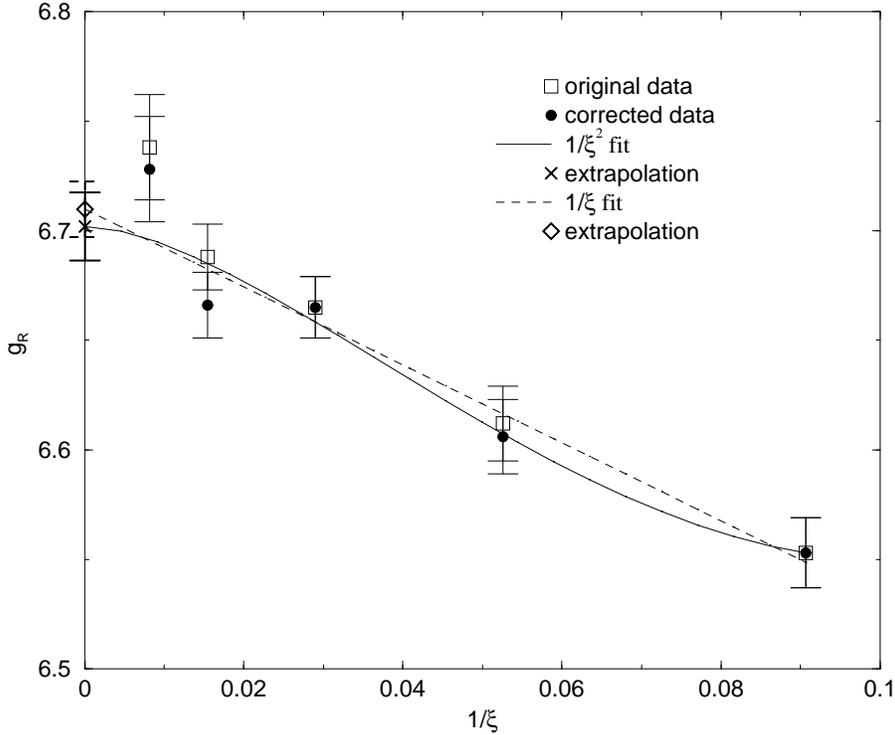,angle=-90,width=12cm}
\caption{\footnotesize The measured values for $\gr$ 
together with the corrected ones at $L/\xi=7.25$ 
for the standard action. The solid and dotted lines correspond
to the fits (\ref{fit1}) and (\ref{fit2}) respectively.}
\label{gr}
\end{figure}

Next we extrapolated the results of our measurements
to the continuum limit.
In Fig.~\ref{gr} we show both the measured and the corrected values of
$\gr$ versus $1/\xi$ for the data 
with $L/\xi\sim7$. It clearly indicates that the
continuum limit of $\gr$ is approached from below as
is the case indicated by some Pad\'{e} analyses and
in the leading order of the $1/n$ approximation
of the lattice theory ~\cite{CPRVgR}.
The only theoretic framework for estimating lattice artifacts comes from
considering Symanzik's~\cite{symanzik} effective action.
The absence of
even parity O(3) invariants with odd engineering dimensions 
suggests an approach to the continuum
limit with leading behavior $(\log\xi)^r/\xi^2$.
As stressed in the introduction, there is no 
rigorous non-perturbative proof of this behavior.

Figure~\ref{gr} shows two fits.
The first is a quadratic fit of the form suggested by 
the Symanzik analysis, where we have taken $r=1$
\footnote{this is the analytic form found in the 
the leading order $1/n$ expansion with 
$b_1=-\frac14,\,b_2=-\frac58\ln 2+\frac14$ ~\cite{CPRVgR}}:
\begin{equation}
  \gr(\xi,z=7.25)=\gr(\infty,z=7.25)\Bigl[ 1+
{b_1\log\xi\over\xi^2} +{b_2\over\xi^2}\Bigr]\,,
\label{fit1}
\end{equation}
with $\gr(\infty,z=7.25)=6.702(16)$, $b_1=-4.4(2.2)$, $b_2=8(5)$.
%Although the fit is reasonable, $\chi^2/{\rm dof}=0.80$,  
%the curve (plotted as a function of $1/\xi$) has
%a vanishing slope at the continuum point (and a negative second
%derivative) which is in conflict with what seems to be suggested
%by the last four data points. 
We made also a second fit in the fig.~\ref{gr} 
of the form:
\begin{equation}
  \gr(\xi,z=7.25)=\gr(\infty,z=7.25)\Bigl[1+{d_2\over\xi}\Bigr]\,,
\label{fit2}
\end{equation}
with $\gr(\infty,z=7.25)=6.710(13)$, $d_2=-0.27(4)$.  
Although there is no
theoretical basis for such a fit, it describes the 
present data as well as the first.
%has a somewhat better $\chi^2/{\rm dof}=0.66$. 

%{\bf Could you please also see how the fits fare when one also include
%the HT results for say $\xi>4??$}

The morale is that the precision of our measurements 
(which are the presently best available) does not
really discriminate between the two fits (or between any intermediate 
fits with leading behavior $(\ln \xi)^r/\xi^2$ with say large $r$).
This unfortunately limits the  
accuracy on the MC determination of the continuum value of $\gr$; 
a conservative estimate would  be $\gr(\infty,z=7.25)=6.71(2)$. 

\begin{figure}[htb]
\begin{center}
%\vspace*{-1mm}
\epsfig{file=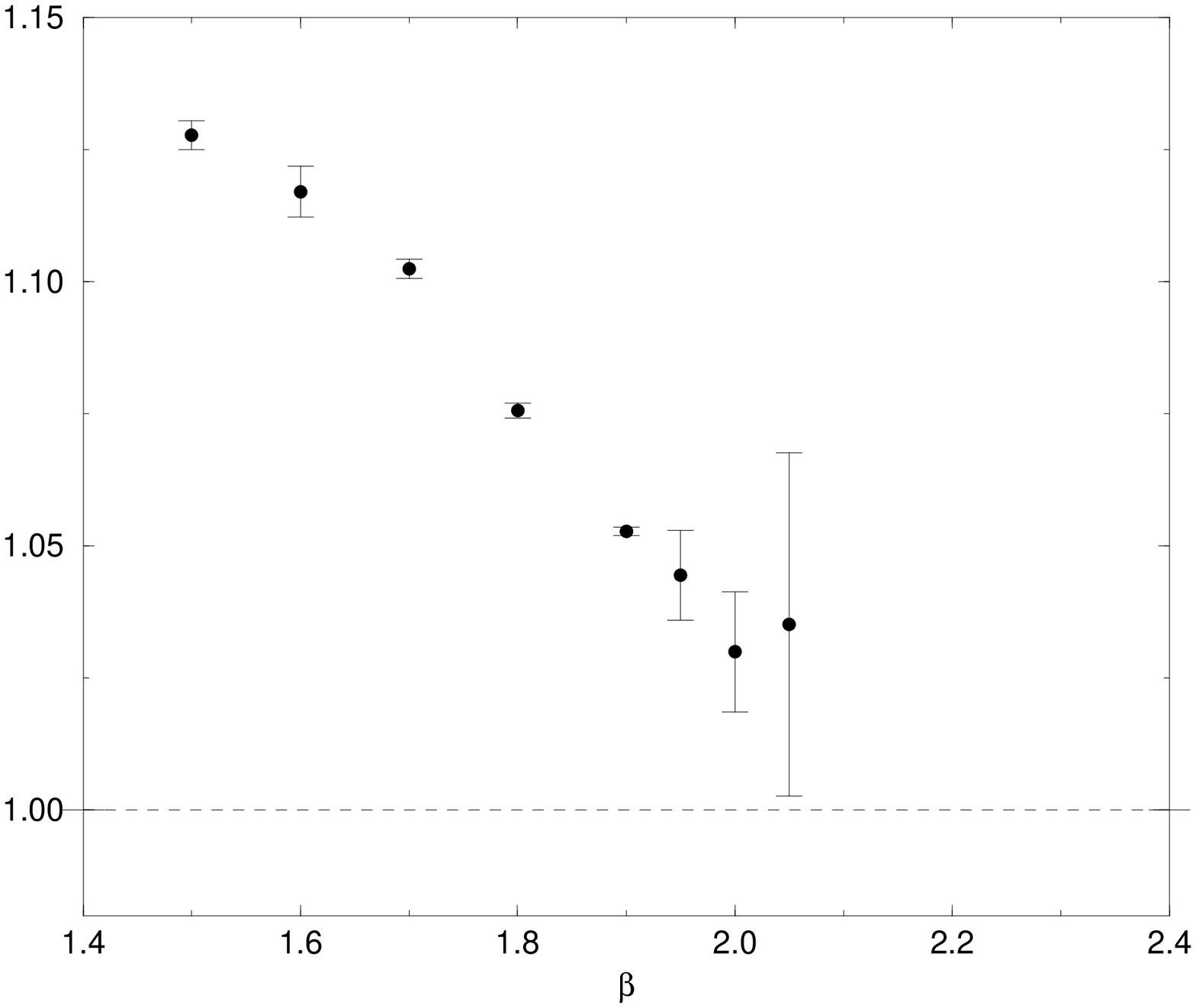,angle=-90,width=12cm}
%\vspace*{-10mm}
\end{center}
\caption{{} \footnotesize
The measured values of the $\chi/\xi^2$ ratio divided by the
4-loop approximation to the prediction (\ref{chioverxi2}).
}
\label{ratio}
\end{figure}

Finally, we used (\ref{FS1n}) again to extrapolate
the continuum result for the finite physical size $z=7.25$ to
the thermodynamic limit $z=\infty$. Thus our final result based on
the standard action lattice measurements is:
\begin{equation}
\gr^{{\rm ST}}=6.77(2)\,.
\label{STRES}
\end{equation}
This error contains both the ambiguity in using the fits (\ref{fit1})
or (\ref{fit2}) and the error in $a_1$.

%\new
%
%\begin{table}[t]
%\centering
%\begin{tabular}[t]{r|r|r|r|r|r}
%\hline
%
%$\beta$ & $L$ &\# of runs & $\xi$\phantom{xxx}& $\chi$\phantom{xxxx}& 
%$\gr$\phantom{xxx}\\[0.5ex]
%\hline \hline
%
%1.5 & $  60$ & $110$\phantom{xx}  & 10.97(2)  & 173.76(31)   &
%6.269(20)\\[0.5ex]
%%\hline
%1.5 & $ 80 $ & $344$\phantom{xx}  & 11.03(1)  & 175.95(11)   &
%6.553(16)\\[0.5ex]
%%\hline
%1.5 & $140 $ & $361$\phantom{xx}  & 11.04(0)  & 176.30(5)    &
%6.560(18)\\[0.5ex]
%%\hline
%1.6 & $140 $ & $214$\phantom{xx}  & 19.02(4)  & 447.30(6)    &
%6.612(17)\\[0.5ex]
%%\hline
%1.7 & $250 $ & $367$\phantom{xx}  & 34.50(2)  & 1267.20(57)  &
%6.665(14)\\[0.5ex]
%%\hline
%1.8 & $500 $ & $361$\phantom{xx}  & 64.79(3)  & 3839.07(1.54)&
%6.688(15)\\[0.5ex]
%%\hline
%1.9 & $910 $ & $101$\phantom{xx}  & 122.32(1) & 11884.9(7.0) &
%6.738(24)\\[0.5ex]
%%\hline
%1.95& $1230$ & $  9$\phantom{xx}  & 167.55(54) & 20882 (67) &
%6.804(92)\\[0.5ex]
%%\hline
%2.0 & $1600$ & $  3$\phantom{xx}  & 229.55(59) & 36694(78) &
%6.9  (6)\\[0.5ex]
%
%\hline
%\end{tabular}
%\caption{\footnotesize Data for $\xi$, $\chi=3G(0)$ and $\gr$}
%\label{grdata}
%\end{table}

\begin{table}[t]
\centering
\begin{tabular}[t]{r|r|r|r|r|r|r}
\hline

$\beta$ & $L$ & runs\phantom{xx} & $L/\xi$\phantom{xx}  & 
$\xi$\phantom{xxx}& $\chi$\phantom{xxxx}& 
$\gr$\phantom{xxx}\\[0.5ex]
\hline \hline

1.5 & $  60$ & 110$\times$20k  & 5.469(10)&
10.97(2)  & 173.76(31)   & 6.269(20)\\[0.5ex]
%\hline
1.5 & $ 80 $ & 344$\times$20k& 7.253(7)\p&
11.03(1)  & 175.95(11)   & 6.553(16)\\[0.5ex]
%\hline
1.5 & $ 100 $ & 350$\times$20k& 9.050(8)\p&
11.05(1)  & 176.51(6)   & 6.613(17)\\[0.5ex]
%\hline
1.5 & $140 $ & 361$\times$20k  & 12.68(1)\p& 
11.04(1)  & 176.30(5)    & 6.560(18)\\[0.5ex]
%\hline
1.6 & $140 $ & 214$\times$20k  & 7.361(15)&
19.02(4)  & 447.30(6)    & 6.612(17)\\[0.5ex]
%\hline
1.7 & $250 $ & 367$\times$20k& 7.246(4)\p&
34.50(2)  & 1267.20(57)  & 6.665(14)\\[0.5ex]
%\hline
1.8 & $500 $ & 361$\times$20k & 7.717(4)\p&
64.79(3)  & 3839.07(1.54)& 6.688(15)\\[0.5ex]
%\hline
1.9 & $910 $ & 101$\times$20k& 7.4395(6)&
122.32(1) & 11884.9(7.0) & 6.738(24)\\[0.5ex]
%\hline
1.95& $1230$ &   12$\times$20k& 7.352(20)&
167.30(45) & 20835 (57) & 6.751(78)\\[0.5ex]
%\hline
2.0 & $1600$ &   4$\times$20k & 6.949(25)&
230.26(83) & 36826(142) & 6.981(132)\\[0.5ex]
%\hline
2.05 & $2100$ &   5$\times$20k & 6.804(67)&
308.63(3.06) & 63011(731) & 6.608(216)\\[0.5ex]
\hline
\end{tabular}
\caption{\footnotesize $\xi$, $\chi=3G(0)$ and $\gr$
for the standard action}
\label{grdata}
\end{table}

% End of passage by Adrian and Erhard
%%%%%%%%%%%%%%%%%%%%%%%%%%%%%%%%%%%%%%%%%%%%%%%%%%%%%%%%%%%%%%%%%%

%\clearpage

%\section{$g_R$ for the FP action}

% (Note that the $\sqrt{z}$ factor is not essential here -- 
% it varies too slowly compared to $\exp(-z)$.)
The data for the FP action seem to lie on a universal curve
(the slope of which roughly corresponds to the $1/n$ prediction)
in spite of the extremely short correlation length.
We interpret this as an indication that the lattice
artifacts for the FP action are small, in any case smaller than 
our error bars.
The measured values for the standard action
show a considerable lattice artifact 
but the data at $z \approx 7$ seem to converge to the FP result 
with increasing $\xi$. Extrapolating the FS effects to $z=\infty$
we get
\begin{equation}
\gr^{{\rm FP}}=6.77(2)
\label{grRESULT}
\end{equation}
and $a_1=5.0(3)$. 
This extrapolation is based on the four $\beta_{\rm FP}=0.85$
data points but including all points of table \ref{fpgr_dat}
does not alter the extrapolation significantly.

Our results (\ref{STRES}) and (\ref{grRESULT}) are
clearly above the value determined
previously via finite size scaling~\cite{kim}, although statistically
compatible with Kim's error estimate. They are
above the central values quoted on the basis of high temperature
expansions discussed above but very consistent with each other and 
the result eq.(\ref{grffb}) from the form factor bootstrap.

% NEW ADDITION OF JANOS

We would like to end this section with a comparison of
an analytic prediction of the ratio $\chi/\xi^2$ with the
MC data. In ref.~\cite{JanosMax}
the perturbative short distance expansion of the spin 2-point function 
was refined to
\begin{equation}
S^{a_1a_2}(x,0)={Z\delta^{a_1a_2}\over3\pi^3}\Big(\ln M\vert x\vert\Big)^2
\Bigg\{1+O\Bigg({\ln\vert\ln M\vert x\vert\vert
\over\ln M\vert x\vert}\Bigg)\Bigg\}\,.
\label{shortdis}
\end{equation}
%was derived (in an admittedly non-rigorous way). 
The new result was the exact (though non-rigorous) determination of the
overall nonperturbative constant.
Using this we can straightforwardly derive the relation \cite{CPRVU}
\begin{equation}
{\chi\over\xi^2}={3\pi\gamma_2^2\over4\delta_2}\,{1\over\beta^2}\,\Bigg\{
1+\sum_{n=1}^\infty\,{c_n\over\beta^n}\Bigg\}\,.
\label{chioverxi2}
\end{equation}
The first three non-universal perturbative coefficients are known 
for the standard action lattice regularization \cite{CaraPeli}:
\begin{equation}
c_1=0.1816\,,\,\,\,c_2=0.1330\,,\,\,\,c_3=0.1362\,.
\end{equation}
Fig. \ref{ratio} shows the measured values of the $\chi/\xi^2$ ratio
divided by the 4-loop approximation to the prediction
(\ref{chioverxi2}).
%Using the $\chi$ and $\xi$ ratio values given in Table~\ref{grdata}
%we find that the ratio of the MC result to 
%the theoretical prediction in the 4-loop approximation is
%\begin{equation}
%\{1.128,1.117,1.102,1.076,1.053,1.030\}
%\label{numbers}
%\end{equation}
%for $\beta=1.5,1.6,1.7,1.8,1.9,2.0$ respectively. 
%
\def\msbar{{\rm \overline{MS\kern-0.05em}\kern0.05em}}
Note that (\ref{chioverxi2}) seems to be satisfied by our data
to a better accuracy than the analogous one for $\xi$ alone \cite{CaraPeli},
but the ratios decrease rapidly and we do not know if
they eventually overshoot the asymptotic value of 1.
Note also that the prediction (\ref{chioverxi2}) is independent of the
$M/\Lambda_{\msbar}$ ratio $8/{\rm e}$ of ref. \cite{HMN}.

%{\bf The notation $c_i$ should be changed either in the above
%formula or in the fit formulae to avoid confusion}

\begin{table}[h]
  \begin{center}
\begin{tabular}{l|l|l|l|l|l|l}
\hline
~$\beta$ & ~$L$ & \quad runs & 
 \quad $L/\xi$  &\p\p\p\p$\xi$ & \qquad $\chi$ & \quad $\gr$ \\[0.5ex]
\hline
\hline
0.70 & 18 & 115$\times$360k& 5.682(3) &\p 3.168(1)&\p 
19.501(8)  &6.493(10) \\[0.5ex]
0.70 & 22 & 197$\times$360k  & 6.924(3) &\p 3.177(1)&\p 
19.646(6)  & 6.674(12) \\[0.5ex]
0.70 & 26 & 363$\times$360k  & 8.176(3) &\p 3.180(1)&\p 
19.686(4)  & 6.761(12) \\[0.5ex]
0.85 & 32 & \p 31$\times$600k& 5.359(5) &\p 5.971(5)&\p 
55.74(5) & 6.417(14) \\[0.5ex]
0.85 & 34 & 189$\times$360k& 5.678(2) &\p 5.988(2)&\p 
56.03(2) & 6.485(8) \\[0.5ex]
0.85 & 36 & \p 52$\times$360k& 6.006(4) &\p 5.994(4)&\p 
56.24(3) & 6.581(13) \\[0.5ex]
0.85 & 42 & 140$\times$360k  & 6.986(3) &\p 6.012(3)&\p 
56.51(2) & 6.691(14) \\[0.5ex]
1.00 & 70 & 691$\times$100k  & 5.734(4) &12.207(8)\p &  
189.4(1)    & 6.487(14) \\[0.5ex] 
\hline
\end{tabular}
    \caption{{} \footnotesize Data for the FP action.
}
    \label{fpgr_dat}
  \end{center}
\end{table}

\clearpage

%\vfill
%\eject

\section{Phase shift analysis from 4-spin correlators}

The prediction for the scattering amplitude
of two particles at center of mass momentum 
$p=M\sinh \frac12 \theta$
in the O(3) non-linear 
sigma model by Zamolodchikov and Zamolodchikov \cite{ZamI} 
is given by
\begin{equation}
S_{a^{\prime}b^{\prime};ab}(\theta)=
\sum_{I=0}^2 e^{2i\delta_I(p)}P^I_{a^{\prime}b^{\prime};ab}\,,
\end{equation} 
where $P^I$ are the isospin projectors and the phase shifts
$\delta_I$ are given simply by 
\begin{gather}
e^{2i\delta_0(p)}={\theta+2i\pi\over \theta-2i\pi}\,,
\\
e^{2i\delta_1(p)}={\theta+2i\pi\over \theta-2i\pi}\,\cdot\,
{\theta-i\pi\over \theta+i\pi}\,,
\\
e^{2i\delta_2(p)}={\theta-i\pi\over \theta+i\pi}\,.
\end{gather}

\begin{table}[htbp]
  \begin{center}
\begin{tabular}{|ccccc|}
\hline
lattice & $\beta$ &  $T\times L$  & $m^{-1}$    &  $mL$    \\
\hline
A & 1.54 & $256\times 128$ & 13.632(6) & 9.4 \\
B & 1.40 & $128\times 64$  &  6.883(3) & 9.3 \\
\hline
D & 0.85 & $128\times 64$  &  6.03(1) & 10.0 \\
E & 0.70 & $64\times 32$   &  3.186(15) & 10.6 \\
\hline
\end{tabular}
\caption{\footnotesize
Parameters of simulations in the O(3) model.
On lattices A, B the standard action was used while D and E
denote simulations with the FP action.}
    \label{lattices}
  \end{center}
\end{table}

%\clearpage

\begin{table}[htbp]
  \begin{center}
\begin{tabular}{|cccccccc|}
\hline
$I$ & $k$ &  $\delta_{\rm exact}$  & $\delta_{\rm E}$    & 
$\delta_{\rm WF}$     & $t_0$ &$M$ & $\delta_{\rm LW}$\\
\hline
0 & 1 & 1.4595    & 1.36(3)    & 1.49(2)    & 1 & 20  & 1.36(7) \\  
  &   &           & 1.45(5)    & 1.50(4)    & 8 &  8  &  \\
  & 2 & 1.2840    & 1.25(2)    & 1.30(2)    & 1 & 20  & 1.22(3) \\
  &   &           & 1.38(6)    & 1.34(5)    & 8 &  8  &  \\
1 & 1 & 0.1751    & 0.10(1)    & 0.17(3)    & 1 & 20  & 0.15(1)\\
  &   &           & 0.19(3)    & 0.19(5)    & 8 &  8  &  \\
  & 2 & 0.2692    & 0.15(2)    & 0.23(2)    & 1 & 20  & 0.23(1) \\
  &   &           & 0.23(4)    & 0.24(5)    & 8 &  8  &  \\
2 & 1 & $-1.3874$ & $-1.51(2)$ & $-1.37(1)$ & 1 & 20  & $-1.39(2)$\\
  &   &           & $-1.43(3)$ & $-1.36(2)$ & 8 &  8  &  \\
  & 2 & $-1.0944$ & $-1.11(1)$ & $-1.06(1)$ & 1 & 20  & $-1.05(1)$\\
  &   &           & $-1.04(2)$ & $-1.06(2)$ & 8 &  8  &  \\
\hline
\end{tabular}
\caption{\footnotesize
Phase shifts from lattice A}   
    \label{resA}
  \end{center}
\end{table}

\begin{table}[htbp]
  \begin{center}
\begin{tabular}{|cccccccc|}
\hline
$I$ & $k$ &  $\delta_{\rm exact}$  & $\delta_{\rm E}$    &
$\delta_{\rm WF}$     & $t_0$ &$M$ & $\delta_{\rm LW}$\\
\hline
0 & 1 & 1.4584    & 1.41(1)    & 1.48(1)     & 1 & 20  & 1.51(7) \\
  &   &           & 1.47(1)    & 1.49(1)     & 3 & 10  & \\
% &   &           & 1.49(2)    & 1.49(1)     & 4 & 10  & \\
  & 2 & 1.2818    & 1.29(1)    & 1.29(1)     & 1 & 20  & 1.2(1)\\
  &   &           & 1.35(1)    & 1.30(1)     & 3 & 10  & \\
% &   &           & 1.38(2)    & 1.31(2)     & 4 & 10  & \\
1 & 1 & 0.1764    & 0.09(1)    & 0.12(1)     & 1 & 20  & 0.14(1)\\
  &   &           & 0.12(1)    & 0.12(1)     & 3 & 10  & \\
% &   &           & 0.12(1)    & 0.11(2)     & 4 & 10  & \\
2 & 1 & $-1.3859$ & $-1.42(1)$ & $-1.35(1)$  & 1 & 20  & $-1.38(2)$\\
  &   &           & $-1.36(1)$ & $-1.35(1)$  & 3 & 10  & \\
% &   &           & $-1.36(1)$ & $-1.35(1)$  & 4 & 10  & \\
  & 2 & $-1.0914$ & $-1.05(1)$ & $-1.03(1)$  & 1 & 20  & $-1.02(1)$\\
  &   &           & $-1.02(1)$ & $-1.03(1)$  & 3 & 10  & \\
% &   &           & $-1.02(1)$ & $-1.03(1)$  & 4 & 10  & \\
  & 3 & $-0.9102$ & $-0.78(2)$ & $-0.78(1)$  & 1 & 20  & $-0.81(2)$ \\
  &   &           & $-0.76(1)$ & $-0.78(1)$  & 3 & 10  & \\
% &   &           & $-0.75(2)$ & $-0.78(2)$  & 4 & 10  & \\
\hline
\end{tabular}
\caption{\footnotesize
Phase shifts from lattice B}
    \label{resB}
  \end{center}
\end{table}

%%%%%%%%%%%%%%%%%%%%%%%%%%%%%%%%%%%%%%%%%%%%%%%%%%%%%%%%%%%%%%%%%

\begin{table}[htbp]
  \begin{center}
\begin{tabular}{|ccccc|}
\hline
$I$ & $k$ &  $\delta_{\rm exact}$  & $\delta_{\rm E}$    &  
$\delta_{\rm WF}$      \\
\hline
0 & 1 & 1.4726    & 1.47(3)    & 1.46(1)    \\
  & 2 & 1.3107    & 1.29(2)    & 1.29(2)    \\
1 & 1 & 0.1598    & 0.16(1)    & 0.14(2)    \\
  & 2 & 0.2545    & 0.23(2)    & 0.27(1)    \\
2 & 1 & $-1.4056$ & $-1.41(2)$ & $-1.41(1)$ \\
  & 2 & $-1.1321$ & $-1.15(1)$ & $-1.16(1)$ \\
\hline
\end{tabular}
\caption{\footnotesize
{}Phase shifts from lattice D.}
    \label{resD}
  \end{center}
\end{table}

\begin{table}[htbp]
  \begin{center}
\begin{tabular}{|ccccc|}
\hline
$I$ & $k$ &  $\delta_{\rm exact}$  & $\delta_{\rm E}$    &  
$\delta_{\rm WF}$     \\
\hline
0 & 1 & 1.4664    & 1.49(1)    & 1.48(1)    \\
  & 2 & 1.2977    & 1.27(1)    & 1.28(1)    \\
1 & 1 & 0.1674    & 0.18(1)    & 0.15(1)    \\
  & 2 & 0.2619    & 0.26(1)    & 0.28(1)    \\
2 & 1 & $-1.3968$ & $-1.35(1)$ & $-1.39(1)$ \\
  & 2 & $-1.1138$ & $-1.14(1)$ & $-1.13(1)$ \\
\hline
\end{tabular}
\caption{\footnotesize
{}Phase shifts from lattice E.}
    \label{resE}
  \end{center}
\end{table}

\begin{figure}[t]
\begin{center}
%\vspace*{-1mm}
\epsfig{file=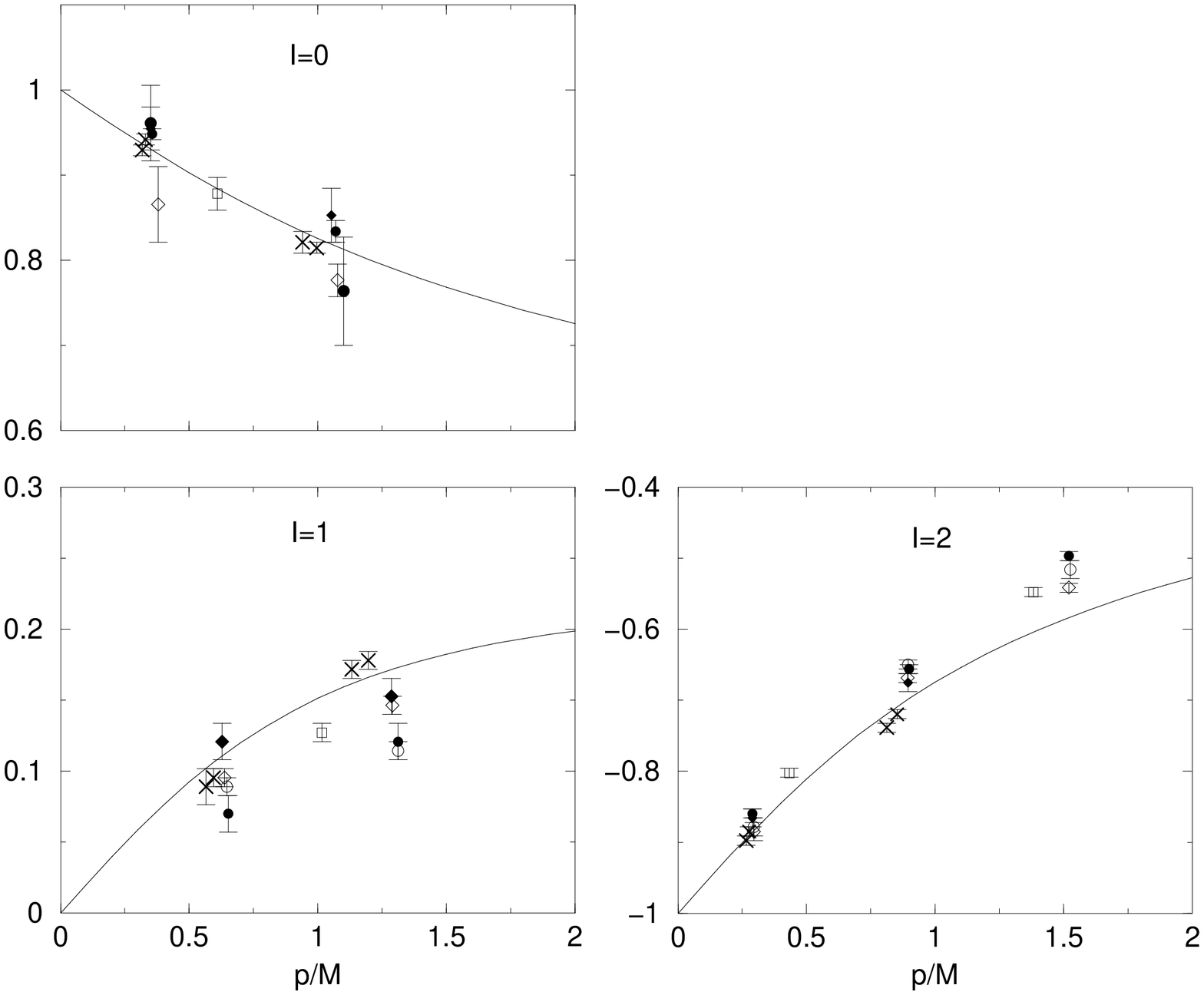,angle=-90,width=12cm}
%\vspace*{-10mm}
\end{center}
\caption{\footnotesize
{}The phase shifts in units of $\pi/2$ vs. $p/M$. 
The open symbols are from
ref.~{\protect \cite{lw}}.
The corresponding filled symbols are our measurements
on the same lattices A and B, with the method discussed here.
The crosses denote results using the FP action (D and E).
The continuous curve corresponds to the ZZ S-matrix.
}
\label{phaseall}
\end{figure}

In fact the Zamolodchikovs specified the S-matrix
for general $n\ge3$ and these expressions have been verified
to O$(1/n^2)$ in the $1/n$-expansion. For the particular
case of O(3), L\"uscher and Wolff \cite{lw}
checked the expressions for low energies using MC simulation.
The agreement was completely satisfactory; 
in particular the data were consistent with
the highly non-trivial non-perturbative property that the S-matrix at
zero energy is repulsive
$S_{a^{\prime}b^{\prime};ab}(0)=-\delta_{a^{\prime}b}\delta_{b^{\prime}a}$,
which is a crucial condition in the FFB construction.
%
%
%As mentioned in the introduction, our initial aim
%to measure the phase shifts was to 
%check some programs and methods of analysis  
%which we are using for the analysis of the xy-model.
%We were hoping that for the standard action we would obtain
%more precise results which would more clearly allow us to see
%the trend of the lattice artifacts. Unfortunately we did
%not achieve this; but it turns out that for the
%fixed point action proposed in ref.~\cite{FP}
%the indications are that the lattice artifacts 
%are much smaller. 
%
%
We repeated the measurements of \cite{lw} for the standard action
using a modified method of analysis. In addition we also
performed measurements using the FP action.
Since the S-matrix is the essential ingredient in the FFB construction
we report the results here.
%
%Since the S-matrix is the essential
%ingredient in the FFB construction we decided
%to report the results here.

The method of L\"uscher and Wolff \cite{lw} is based on
the following idea:
The momentum of one of the particles in a 2-particle state 
with zero total momentum in a periodic box (in 1-dimension) 
takes discrete values $p_n$ given by the periodicity condition
\begin{equation}
  \label{delta}
  p_n L + 2 \delta(p_n) = 2\pi n \,. 
\end{equation}
Accordingly, the energy of this state is given by
\begin{equation}
  \label{En}
  E_n = 2 E^{(1)}(p_n)= 2 \sqrt{p_n^2 + M^2} \,,
\end{equation}
where $E^{(1)}(p)$ is the energy of a 1-particle state with momentum $p$.
From the measurement of the energy spectrum 
$E_n$ for some low lying states one can then calculate the   
momentum $p_n$ and using eq.~(\ref{delta}) the phase shift $\delta(p_n)$.
Varying $L$ and taking different values of $n$ one can determine
$\delta(p)$ at several values of its argument.

To determine the 2-particle energies the correlation matrix has been 
measured\footnote{Actually, in Ref.\cite{lw} the measurement was done in 
Fourier space (i.e. in relative momenta), but this difference is not 
significant here. For our purpose the coordinate space representation 
is more convenient.}
\begin{equation}
  \label{Ct}
  C_{xy}(t)=\langle {\rm vac}| O(x,0) O(y,t) |  {\rm vac} \rangle_{\rm c}
%  -\langle {\rm vac}| O_x | {\rm vac} \rangle 
%   \langle {\rm vac}| O_y | {\rm vac} \rangle 
\,,
\end{equation}
where
\begin{equation}
  \label{Ox}
  O(x,t)=\frac{1}{L} \sum_{z=0}^{L-1} \sigma(z,t)\sigma(z+x,t) \,.
\end{equation}
We omit here the $O(n)$ structure and indices. The subscript c in 
eq.~(\ref{Ct}) means that in the $I=0$ channel the vacuum contribution 
is subtracted.

Using the eigenvectors $|n\rangle$ of the transfer matrix as
intermediate states one has
\begin{equation}
  \label{Cxy}
  C_{xy}(t)= \sum_n {\rm e}^{-E_n t} \psi_n(x) \psi_n(y)  \,,
\end{equation}
where
\begin{equation}
  \label{psix}
  \psi_n(x)= \langle {\rm vac} | O(x,0) |n\rangle 
\end{equation}
is the ``wave function'' of the corresponding state.
The lowest energy states in eq.~(\ref{Cxy}) are the 2-particle states
and they will dominate at sufficiently large values of $t$. 

Note that the relative momentum $2p_n$ of the 2-particle states is encoded 
not only in the energy $E_n$ but also in the wave 
function $\psi_n(x)$. For the symmetric wave functions 
($I=0,2$ channels) one should have
\begin{equation}
  \label{psi0}
  \psi_n(x)= A \cdot\cos p_n(x-L/2), \text{   for } R < x < L-R \,,
\end{equation}
and similarly  with  $\sin p_n(x-L/2)$ for the $I=1$ channel.
Here $R$ is the `interaction range': for a relative distance
$x>R$ the particles propagate (essentially) freely. 

One expects that the additional information obtained from
the wave function will provide a more precise determination of $p$
and hence of $\delta(p)$.

\subsection{Determination of the energy spectrum and wave functions.}

The rank $N$ of the matrix $C(t)$ in eq.~(\ref{Cxy}) is
$L/2$, $L/2-1$ and $L/2+1$ in the $I=0,1,2$ channels, respectively.
We assume that for $t \ge t_0$ (with some $t_0$) no more than $N$
states contribute to $C_{xy}(t)$, 
in the sense that the total contribution of the
states $n>N$ is much smaller than the statistical error 
$\delta C_{xy}(t)$.

L\"uscher and Wolff \cite{lw} 
suggested to determine the energies
of the 2-particle states from the generalized eigenvalue problem
\footnote{This equation was considered already before
by Michael ~\cite{Michael}, in connection with a variational
approach for evaluating the static potential in lattice
gauge theory.}
\begin{equation}
  \label{genev}
 C(t) v_n = \lambda_n(t,t_0) C(t_0) v_n \,.
\end{equation}

The eigenvalues of eq.~(\ref{genev}) are given {\em exactly} by
\begin{equation}
  \label{lambdan}
 \lambda_n(t,t_0) = {\rm e}^{-E_n(t-t_0)} \,,
\end{equation}
provided the sum in eq.~(\ref{Cxy}) is restricted to $N$ terms, 
$1\le n \le N$.
It is also easy to show that (with an appropriate normalization of 
$v_n$):
\begin{equation}
  \label{psin}
 \psi_n(x)=\sum_y C_{xy}(t_0) v_n(y) \,.
\end{equation}

Solving eq.~(\ref{genev}) involves an inversion of $C(t_0)$,
and the distortion of its small eigenvalues by the statistical
noise is enhanced. This could affect strongly the values and the
errors of $E_n$ obtained. 
For $t_0>1$ taking all $N \sim L/2$ states introduces significant
instability in the result.\footnote{In ref.\cite{lw} $\sim L/4$
states were used with $t_0\le 1$.}
Because of this, we have introduced a modification:
before considering the generalized eigenvalue problem, we truncate
the correlation function to an $M$-dimensional subspace ($M<N$)
spanned by the first $M$ eigenvectors of $C(t_0)$ (to those with
the largest $M$ eigenvalues and still stable against the statistical
fluctuations). The generalized eigenvalue problem, eq.~(\ref{genev}) 
is written then for the matrices $\overline{C}(t)$ in this reduced basis.
The energies we used were obtained from eqs.~(\ref{genev},\ref{lambdan}).
We also determined $v_n(x)$ from eq.~(\ref{genev}).
One can use them as some given (nearly optimal) projectors
which satisfy
\begin{equation}
  \label{vCv}
\left( v_m , C(t) v_n \right) = \delta_{mn}  {\rm e}^{-E_n(t-t_0)} \,.
\end{equation}
Note that due to statistical errors in $C_{xy}(t)$ the obtained 
$v_n(x)$'s will differ from the true ones hence this equations 
will be only approximately valid. They provide a useful consistency
check, and also a somewhat different method to determine $E_n$.

As an alternative way to get the phase shifts we used the momenta 
$p_n$ determined from the wave function using eq.~(\ref{psin}).
For a given $t_0$ and $M$ one can also check the self consistency of 
the obtained parameters by comparing $C_{xy}(t)$ built from $E_n$ and
$\psi_n(x)$ (cf. eqs.~(\ref{lambdan},\ref{psin})) with the MC result.

\subsection{The results}

We performed the calculations with the standard action for the same 
lattices (A,B) as in Ref.~\cite{lw}
\footnote{Ref.~\cite{lw} also measured an additional lattice C
with $mL\sim 5$.}.  
In addition, we also repeated the calculations using the fixed point 
action (FP) for the O(3) model \cite{FP,FP1}.
The parameters of our measurements are summarized in Table~\ref{lattices}
(here $m$ is the inverse of the exponential correlation length).

The phase shifts obtained from the analysis for the standard 
and FP actions are shown in fig.~\ref{phaseall}
together with the results from ref.~\cite{lw}.
The results for the standard action are also given in tables 
\ref{resA} and \ref{resB} and for the FP action in tables
\ref{resD} and \ref{resE}.
The column $\delta_{\rm E}$ gives the phase shift calculated 
from the energy by eq.~(\ref{genev}), WF labels the results 
obtained from the wave function, eq.~(\ref{psin}). 
The data shown correspond to $t_0=3$, $M=10$ for lattice D,
and $t_0=1$, $M=10$ for lattice E. 
However, the results -- especially for $\delta_{\rm WF}$ --
are quite stable against this choice.
We took $R \sim 3/m$ in eq.~(\ref{psi0}).
Taking into account that the correlation lengths used with the FP
action were only $\xi \approx 3$ and $6$, the agreement with the analytic
prediction is very good. Note, however, that similar suppression
of lattice artifacts has been observed previously with this action
for other observables \cite{FP,FP1,Spiegel}.

%A more detailed description of the analysis used will be given
%in ref.~\cite{BNNPSWIII}.

%%%%%%%%%%%%%%%%%%%%%%%%%%%%%%%%%%%%%%%%%%%%%%%%%%%%%%%%%%%%%%%%%%

\clearpage
%\vfill
%\eject

\noindent{\it Acknowledgements}

We would like to thank Michael Karowski for useful discussions, and 
Paolo Butera, Paolo Rossi and Ettore Vicari for informative
correspondence.

This investigation was supported in part by the Hungarian National
Science Fund (OTKA) (under T016233 and T019917), and also
by the Schweizerischer Nationalfonds.
The work of M.N. was supported by NSF grant 97-22097.

%\clearpage

%\vfill 
%\eject

% List of references

\end{document}